\documentclass[12pt]{iopart}
\newcommand{\be}{\begin{equation}}
\newcommand{\ee}{\end{equation}}
\newcommand{\dg}{^{\dagger}}
\newcommand{\frp}[2]{\frac{\partial#1}{\partial#2}}
\newcommand{\frd}[2]{\frac{\de#1}{\de#2}}
\newcommand{\de}{{\rm d}}
\newcommand{\bibt}[1]{{\it #1}}
\newcommand{\arch}[1]{#1}
\begin{document}

\title{Flow Equations and Normal Ordering. A Survey}
\author{Franz Wegner}
\address{Institut f\"ur Theoretische Physik, Universit\"at Heidelberg,
Philosophenweg 19, D-69120 Heidelberg}

\begin{abstract}
First we give an introduction to the method of diagonalizing or
block-diagonalizing continuously a Hamiltonian and explain how this procedure
can be used to analyze the two-dimensional Hubbard model. Then we give a short
survey on applications of this flow equation on other models. Finally we
outline, how symmetry breaking can be introduced by means of a symmetry breaking
of the normal ordering, not of the Hamiltonian.
\end{abstract}

\section{Introduction}

Renormalization group plays an important role both in high-energy physics and in
condensed matter physics. It was invented in high-energy
physics\cite{Stueckelberg53,Gell-Mann54} in order to
resolve the problems with divergencies at large momenta. It was developed as a
tool to define how an interaction with cut-off at some momentum $\Lambda$ has
to vary as $\Lambda$ increases {\it up to} the limit infinity so that the
expectation values converge to a finite limit.

For a long time the explanation of critical behavior was an open question in
condensed matter physics. Kenneth Wilson\cite{Wilson71} showed along ideas
developed by Leo Kadanoff \cite{Kadanoff65} that also this problem is
solved by the renormalization group. In condensed matter most often one does
not worry about large momenta, in particular if one considers a system on a
lattice where momenta are restricted to the Brillouin zone. Then the problem of
critical phenomena resides in small momenta and one has to integrate out the
contributions of the interaction at momenta {\it down to} some small cut-off
$\Lambda$ and finally to consider the limit where this cut-off approaches zero.
This applies for bosonic degrees of freedom which have the advantage that their
quantum nature is normally irrelevant so that they can be considered as
classical fields.

The situation is different for fermionic degrees of freedom. They do not show a
classical limit. Therefore other methods had to be developed.
The essential physics in a condensed fermionic system comes from the behaviour
at the Fermi edge. Therefore it is natural to focus on the behaviour close
the Fermi edge and to eliminate the degrees of freedom away from this Fermi
edge so that the cut-off describes now the distance from the Fermi edge. This
procedure was pushed forward by Shankar\cite{Shankar91,Shankar94}. It has been
applied to fermionic systems among others by Zanchi and Schulz\cite{Zanchi00},
Salmhofer and Honerkamp\cite{Salmhofer01}, Halboth and Metzner\cite{Halboth00}

There is a second approach which was started in 1993/94. It appeared under the
name of {\it Flow equations for Hamiltonians}\cite{Glazek93,Glazek94} and {\it
Similarity transformation}\cite{Wegner94} and is also often called {\it
Continuous unitary transformation} [CUT]. G{\l}azek and Wilson and myself were
at
that time unaware that mathematicians in the field of control theory had
developed similar ideas under the
names {\it Double Bracket Flow}\cite{Brockett91} and {\it Isospectral
Flow}\cite{Chu90,Chu94}. The basic idea is
to choose the Hamiltonian in a certain basis, e.g. the basis of Bloch waves and
then to eliminate first the off-diagonal matrix elements between states which
differ strongly in energy. This is done by a continuous unitary transformation
as function of a flow-parameter $\ell$. We start from the initial
Hamiltonian $H=H(0)$ and obtain a Hamiltonian $H(\ell)$ by means of a unitary
transformation $U(\ell)$,
\be
H(\ell) = U(\ell) H U\dg(\ell).
\ee
Differentiation with respect to $\ell$ yields
\be
\frd{H(\ell)}{\ell} = [\eta(\ell),H(\ell)]
\ee
with the generator $\eta$ of the unitary transformation
\be
\eta(\ell) = \frd{U(\ell)}{\ell} U\dg(\ell) = -\eta\dg(\ell).
\ee
This is the flow equation for the Hamiltonian. Obviously $U(\ell)$ or actually
$\eta(\ell)$ have to be chosen in an appropriate way. There are various ways of
doing so.

\section{Various Choices for the Generator}

\newcommand{\Hd}{H^{\rm d}}
\newcommand{\Hr}{H^{\rm r}}
The most obvious way is to choose a generator which nearly always diagonalizes
the Hamiltonian. If one chooses
\be
\eta=[\Hd,H],
\ee
where $\Hd$ is the diagonal part of the Hamiltonian, then we find that the sum
of the squares of the off-diagonal matrix elements decays like
\be
\sum_{k,k'|k\not=k'} \frp{h_{k,k'}h_{k',k}}{\ell}
=-2\sum_{k,k'} (\epsilon_k-\epsilon_{k'})^2 h_{k,k'}h_{k',k},
\ee
where the diagonal matrix elements are denoted by $\epsilon$.\footnote{The flow
equation now reads $\de H/\de\ell=[[\Hd,H],H]$ which explains the notion
double-bracket equation.} Thus the off-diagonal matrix elements decay unless
(which happens only rarely) the elimination stops with a non-zero off-diagonal
element between two degenerate levels.\footnote{These argument as well as the
arguments on the cost function below apply rigorously for finite matrices. For
the infinite dimensional Hilbert space they are guidelines.}
Note however, that in general also $\epsilon$ depends on $\ell$.\footnote{In the
case of the elimination of the electron-phonon coupling these couplings decay
even if the states are finally degenerate.\cite{Lenz96} Depending on the
elimination scheme such off-diagonal matrix elements can survive in the Kondo
Hamiltonian below the Kondo temperature\cite{Thimmel96}}

Although it seems quite desirable to diagonalize the Hamiltonian completely,
there is the draw-back that nearly always approximations have to be introduced.
In order that they do not create too large errors it is advisable to perform
only weak unitary transformations. This can be done by bringing the Hamiltonian
to a block-diagonal form. In the case of the elimination of the electron-phonon
coupling we eliminate only the matrix elements which change the number of the
phonons, that is the electron-phonon coupling. In this case we are left with an
effective electron-electron interaction which has the nice properties (i) that
it is attractive between all pairs of electrons with total momentum zero in
contrast to Fr\"ohlich's Hamiltonian and (ii) it is instantaneous. This second
property is inherent to the scheme of flow equations.

Another case where we bring the Hamiltonian in block-diagonal form is the
elimination of the contributions in an electronic system which does not
conserve the number of quasi-particles (electrons above and holes below the
Fermi edge). Then the eigenstates are states with fixed numbers of these
quasi-particles, so that the excitation energies can be read off the
one-particle contribution of the effective Hamiltonian and the two-particle
excitations are now excitations determined from a two-particle
problem.\cite{Wegner94}

With $H$ consisting of the diagonal and the off-diagonal part $H=\Hd+\Hr$ the
genarator $\eta$ may be rewritten
\be
\eta = [H,\Hr].
\ee
We are free, however, to modify $\Hr$. If we choose
\be
\Hr=[N,[N,H]],
\ee
where $N$ is the particle operator of the phonons or of the quasi-particles,
then the particle number violating terms will be eliminated. To see this we may
introduce a cost function
\be
G(H) = \frac 12 \sum g_{ij,kl} H_{ji} H_{lk}
=\frac 12 \tr (H \Hr), \quad
\Hr_{ij} = \sum g_{ij,kl} H_{lk}
\ee
where we require that $g$ is symmetric and the cost function is real,
\be
g_{ij,kl}=g_{kl,ji}=g^*_{ji,lk}.
\ee
Then one obtains
\be
\frd G{\ell} = \tr([\eta,H] \Hr) = \tr(\eta [H,\Hr]).
\ee
If $G$ is semi-positive definite then the choice
\be
\eta = [H,\Hr]
\ee
yields
\be
\frd G{\ell} = \tr([H,\Hr] [H,\Hr]) \le 0.
\ee
Note that $\eta$ is anti-hermitean. The derivative $\frd G{\ell}$ vanishes only,
if $\Hr$ commutes with $H$. In all other cases the cost function decreases. The
choice $\Hr=[N,[N,H]]$ yields the cost function
\be
G=\frac 12 \tr([H,N] [N,H])
\ee
which becomes zero only if $H$ commutes with $N$.

This allows another procedure useful for systems with symmetry breaking. We may
choose quite generally
\be
\Hr=\sum_{\alpha} [v^{\alpha},[v^{\alpha},H]]
\ee
If $v$ is a one-particle operator
\be
v=\sum_k v_k c\dg_k c_k,
\ee
then the evaluation of $[v,[v,H]]$ multiplies terms of type
$c\dg_{k_1}c\dg_{k_2}... c_{q_1}c_{q_2}... $ in $H$ by
\be
r_{k_1k_2...q_1q_2...} = (v_{k_1}+v_{k_2} +... -v_{q_1}-v_{q_2}-...)^2.
\ee
This elimination function indicates how urgently we wish to eliminate such terms
in the Hamiltonian. We have used this form of $\Hr$ in order to calculate the
effective potential of the Hubbard model described by the Hamiltonian
\be
H= -t \sum_{\rm n.n.} c\dg_{r's} c_{rs} - t' \sum_{\rm n.n.n.} c\dg_{r's} c_{rs}
+ U \sum_r (n_{r\uparrow} - \frac 12) (n_{r\downarrow} - \frac 12 ).
\ee
in second order in the coupling $U$ (weak coupling limit). We used the condition
\be
-v_{-k} = v_k = v_{k+q_0}
\ee
which transformed the Hamiltonian into a molecular-field form for the expected
order parameters (superconductivity, antiferromagnetism, flux-phases,
Pomeranchuk instability) to be expected, that is we allowed for non-zero
\be
\langle c\dg_{ks} c\dg_{-ks'} \rangle, \quad
\langle c\dg_{ks} c_{k+q_0s'} \rangle, \quad
\langle c\dg_{ks} c_{ks'} \rangle \quad \mbox{and even} \quad
\langle c\dg_{ks} c\dg_{-k+q_0s'} \rangle. \label{order}
\ee
Depending on the couplings the above mentioned instabilities showed up. The temperature dependence enters via normal ordering. Since the renormalization flow does not only alter two-particle interactions but also generates higher particle-interactions the two-particle interaction depends on the normal ordering. More over the expansion of the entropy in terms of the expectation values (\ref{order}) yields a temperature dependence.
For more details see \cite{Grote02,Hankevych02,Hankevych03a,Hankevych03b}.

\section{Other Applications}

Meanwhile the method of flow equations has been applied to many systems. Apart
from those already mentioned we list the following applications:
Anderson impurity model\cite{Kehrein94,Kehrein96b,Stauber03b}
Fano-Anderson and Anderson lattice\cite{Becker02},
spin-boson models and
dissipation\cite{Kehrein96a,Kehrein96c,Kehrein97,Kehrein98,Mielke98,Mielke00,Stauber02a,Stauber02b,Stauber02c,Kleff03a},
qbit and spin-boson model\cite{Kleff03b},
electron-phonon interaction and
superconductivity\cite{Mielke97a,Mielke97b,Moca98a,Ragwitz99,Huebsch02,Dusuel05a}, superconductivity and impurities\cite{Crisan97},
boson fermion model\cite{Moca98b,Domanski01,Domanski03a,Domanski03b},
Tomonaga-Luttinger model\cite{Stauber02d,Stauber03a},
one-dimensional fermions\cite{Heidbrink02},
Kondo model\cite{Vogel97,Hofstetter01,Kehrein02,Slezak02,Lobaskin04,Kehrein04,Sommer05,Vogel05},
Fermi and Luttinger liquid\cite{Kabel97,Heidbrink01},
sine-Gordon model\cite{Kehrein99,Kehrein01},
QED\cite{Brisudova96,Jones96,Gubankova98},
QCD and general\cite{Wilson94,Walhout98,Pauli00,Glazek00,Glazek01,Glazek03b,Glazek03c},
two-dimensional $\delta$-potential\cite{Glazek97a,Glazek97b,Szpigel99,Szpigel00,Glazek02b},
limit cycles and three-body problem\cite{Glazek02a,Glazek03a},
mapping of the Hubbard-Model\cite{Stein97,Reischl04},
Heisenberg antiferromagnet\cite{Stein98,Stein00a},
spin-Peierls transition\cite{Uhrig98},
spin models\cite{Knetter00,Raas01,Brenig01,Brenig02,Schmidt02a,Schmidt02b,%
Schmidt03,Knetter03a,Knetter03b,Sykora04,Uhrig05},
RKKY interaction\cite{Stein99},
heavy fermions\cite{Meyer04},
interacting Bosons\cite{Dusuel05b,Reischl05},
Lipkin model\cite{Mielke98,Pirner98,Stein00b,Bartlett03,Scholtz03,Kriel05},
Lipkin-Meshkov-Glick model\cite{Dusuel04b,Dusuel04c},
Dirac particle\cite{Bylev98},
molecules\cite{White02},
Henon-Heiles Hamiltonian\cite{Cremers99},
quartic Oscillator\cite{Dusuel04a},
complex eigenvalues\cite{Ohira02}.
It may be mentioned that the two-beam coupling in photorefractive media itself obeys the flow equation scheme\cite{Anderson99}.

The author gave short reviews on the flow equation method at several occasions
\cite{Wegner98a,Wegner98b,Wegner00a,Wegner00b,Wegner00c} mainly explaining the elimination of the electron-phonon interaction and the application to an $n$-orbital model in the limit of large $n$, based mainly on \cite{Wegner94,Lenz96}.
Stefan Kehrein prepares a book on the flow
equation approach to many-body problems\cite{Kehrein06}.

\section{Symmetry Breaking}

In applying flow equations to a Hamiltonian one typically starts out from a
Hamiltonian which does not show an explicit symmetry breaking even if the
symmetry will be broken below some temperature. 

The same applies for the
renormalization group flows. As indicated above in the case of flow
equations one can bring the effective interaction to molecular-field form and finally apply molecular field theory.
Since we transform the interaction no divergencies appear. In the case of the
fermionic renormalization group the vertex functions will at least within
weak-coupling approximations diverge at some length scale, so that the
approximations become unreliable and one has to resort to other methods in this
regime. Thus it is desirable to have a way to introduce symmetry breaking from
the beginning. Recently Salmhofer, Honerkamp, Metzner, and
Lauscher\cite{Salmhofer05} have added a symmetry breaking field to the
Hamiltonian and showed that this leads into the symmetry broken phase.

For the Hamiltonian flow it is not necessary to add a symmetry breaking term to
the Hamiltonian. Instead it is sufficient to choose a normal-ordering which is
symmetry broken. One can show\cite{Koerding05} that the system will nearly
always converge to
the stable state, that is in case of symmetry breaking (that is below the
critical temperature) it runs to a symmetry broken state, whereas if the
symmetric state is stable (above $T_c$) then it will run to the symmetric state.
The basic idea is the following: Normal ordering is given by the expectation
values of bilinear expectation values
\be
G_{kj} = \langle a_k a_j \rangle,
\ee
where the $a_k$ stand for creation and annihilation operators $c\dg_k$ and
$c_k$.
In general one will have normal and anomal expectation values. For given
expectation values $G$ one obtains from an operator $A$ the normal-ordered one
\be
:A_G:_G = A
\ee
by
\be
A_G = \exp\Big( \sum_{kj} G_{kj} \frp{^2}{a_j^{\rm right} \partial a_k^{\rm
left}} \Big) A(a).
\ee
In this expression the operators $a$ anticommute for fermions. Summation runs
over all pairs $a$, where $a_j$ is to the right of $a_k$.
An infinitesimal change in $G$ leads to a change of $A_G$
\be
\delta A_G = A_{G+\delta G}-A_G = \frac 12 \sum_{kj} \delta G_{kj}
\frp{^2}{a_j \partial a_k} A_G.
\ee
Thus under the Hamiltonian flow the Hamiltonian $H_G$ will change due to the
unitary transformation and due to the change of normal ordering yielding
\be
\frd{H_G}{\ell} = [\eta,H_G] + \frac{\delta H_G}{\delta G} \frp G{\ell}.
\ee
As expectation values $G$ entering the normal ordering we may use the expectation values of $\exp{-\beta H^0}$ with the one-particle Hamiltonian
\be
H^0 = \sum_{kj} \frac 12 \tilde{\epsilon}_{kj} a\dg_k a_j.
\ee
Besides the flow equation for the Hamiltonian a flow equation for $H^0$ has to
be introduced. This can be done by requiring that $\tilde{\epsilon}$ approaches
the one-particle part of $H_G$,
\be
H_G = v^{(0)} + \frac 12 \sum_{kj} v^{(1)}_{k^*j} a_k a_j + O(a^4),
\ee
where $a_{k*}=a\dg_k$ by means of the equation
\be
\frp{\tilde{\epsilon}_{kj}}{\ell}
= \gamma (v^{(1)}_{k^*j}-\tilde{\epsilon}_{kj})
\ee
with some positive constant $\gamma$.

Under the following assumptions one can show\cite{Koerding05} that the flow approaches a stable
phase:
To show this we determine the free energy with respect to the density operator
$\exp(-\beta H^0)/Z^0$. If this free energy
\begin{eqnarray}
F^0= \langle H \rangle_0 - TS, \quad \langle H \rangle_0 = v^{(0)}, \nonumber\\
S= -\frac{k_{\rm B}}2 \tr\Big(\tilde G \ln \tilde G+(1-\tilde G) \ln(1-\tilde
G)\Big), \quad \tilde G_{kj} = G_{k^*j}
\end{eqnarray}
is a local minimum against variation of $H^0$, then the fixed point of
$H(\infty)$ is stable; if it is not a local minimum then it is unstable.
Without performing the flow, that is for $\eta\equiv 0$, this procedure approaches nearly always the Hartree-Fock-Bogoliubov solution.

\end{document}